\documentclass[pdflatex,sn-mathphys]{sn-jnl}

\jyear{2021}%

\theoremstyle{thmstyleone}%
\theoremstyle{thmstyletwo}%

\theoremstyle{thmstylethree}%
\graphicspath{{figures/}} 

\usepackage{derivative}

\usepackage{xr}
\makeatletter

\newcommand*{\addFileDependency}[1]{
\typeout{(#1)}
\@addtofilelist{#1}
\IfFileExists{#1}{}{\typeout{No file #1.}}
}\makeatother

\usepackage{subcaption}
\usepackage[section]{placeins}
\usepackage{tabularx}
\usepackage{multirow}
\usepackage{CJKutf8}
\usepackage{amsmath} 

\usepackage{float}
\usepackage{etoolbox}
\makeatletter
\patchcmd{\ps@headings}
{\hbox to \hsize{\hfill Springer Nature 2021 \LaTeX\ template\hfill}}
{\hbox to \hsize{}}
{}
{}
\patchcmd{\ps@titlepage}
{\hbox to \hsize{\hfill Springer Nature 2021 \LaTeX\ template\hfill}}
{\hbox to \hsize{}}
{}
{}
\makeatother
\makeatletter
\patchcmd{\ps@headings}
{\hbox to \hsize{\hfill Springer Nature 2021 \LaTeX\ template\hfill}}
{\hbox to \hsize{}}
{}
{}
\patchcmd{\ps@headings}
{\hbox to \hsize{\hfill Springer Nature 2021 \LaTeX\ template\hfill}}
{\hbox to \hsize{}}
{}
{}
\patchcmd{\ps@titlepage}
{\hbox to \hsize{\hfill Springer Nature 2021 \LaTeX\ template\hfill}}
{\hbox to \hsize{}}
{}
{}
\makeatother

\begin{document}

\begin{CJK*}{UTF8}{gbsn}

\title[FuXi-2.0]{FuXi-2.0: Advancing machine learning weather forecasting model for practical applications}

\author[1]{\fnm{Xiaohui} \sur{Zhong}}\email{x7zhong@gmail.com}
\equalcont{These authors contributed equally to this work.}
\author[1]{\fnm{Lei} \sur{Chen}}\email{cltpys@163.com}
\equalcont{These authors contributed equally to this work.}

\author[2]{\fnm{Xu} \sur{Fan}}\email{fanxu@sais.com.cn}

\author[1]{\fnm{Wenxu} \sur{Qian}}\email{wxqian24@m.fudan.edu.cn}

\author[1]{\fnm{Jun} \sur{Liu}}\email{liujun$\_$090003@163.com}


\author*[1,2]{\fnm{Hao} \sur{Li}}\email{lihao$\_$lh@fudan.edu.cn}

\affil[1]{\orgdiv{Artificial Intelligence Innovation and Incubation Institute}, \orgname{Fudan University}, \orgaddress{\city{Shanghai}, \postcode{200433}, \country{China}}}

\affil[2]{\orgname{Shanghai Academy of Artificial Intelligence for Science}, \orgaddress{\city{Shanghai}, \postcode{200232}, \country{China}}}



\abstract{
Machine learning (ML) models have become increasingly valuable in weather forecasting, providing forecasts that not only lower computational costs but often match or exceed the accuracy of traditional numerical weather prediction (NWP) models. Despite their potential, ML models typically suffer from limitations such as coarse temporal resolution, typically 6 hours, and a limited set of meteorological variables, limiting their practical applicability. To overcome these challenges, we introduce FuXi-2.0, an advanced ML model that delivers 1-hourly global weather forecasts and includes a comprehensive set of essential meteorological variables, thereby expanding its utility across various sectors like wind and solar energy, aviation, and marine shipping. 
Our study conducts comparative analyses between ML-based 1-hourly forecasts and those from the high-resolution forecast (HRES) of the European Centre for Medium-Range Weather Forecasts (ECMWF) for various practical scenarios. The results demonstrate that FuXi-2.0 consistently outperforms ECMWF HRES in forecasting key meteorological variables relevant to these sectors.
In particular, FuXi-2.0 shows superior performance in wind power forecasting compared to ECMWF HRES, further validating its efficacy as a reliable tool for scenarios demanding precise weather forecasts.
Additionally, FuXi-2.0 also integrates both atmospheric and oceanic components, representing a significant step forward in the development of coupled atmospheric-ocean models. 
Further comparative analyses reveal that FuXi-2.0 provides more accurate forecasts of tropical cyclone intensity than its predecessor, FuXi-1.0, suggesting that there are benefits of an atmosphere-ocean coupled model over atmosphere-only models.
}


\maketitle

\section{Introduction}

Weather forecasting is essential for operational efficiency and safety in sectors like renewable energy, aviation, and marine shipping \cite{wmo2015}. Accurate forecasts of wind speed and solar irradiance are vital for maintaining grid stability and reducing costs associated with integrating renewable energy into power grids \cite{yang2019universal,aslam2021survey,zafar2022adaptive}.
For instance, a 1\% improvement in wind forecast could reduce generation costs by 0.27\% \cite{mc2013value}, while a 25\% enhancement in solar energy forecasts could lower net generation costs by 1.56\% \cite{yagli2019reconciling}.
By the end of 2023, China had installed 441 GW of wind and 609 GW of solar power capacity, as reported by the China's National Energy Administration (NEA), highlighting the growing need for accurate forecasts as renewable energy penetration increases \citep{Foley2012,bird2016wind}. 
In aviation, weather accounts for approximately 75\% of flight delays according to the National Airspace System (NAS) \cite{wolfson2006advanced}, making accurate weather forecasting essential for flight planning, navigation, and safety.
Better forecasts allow airlines to optimize fuel consumption and mitigate the effects of adverse weather conditions \cite{anaman2017benefits}.
In marine shipping, which transports over 80\% of global merchandise by volume and contributes nearly 3\% of global greenhouse gas (GHG) emissions \cite{UNCTAD2023}, accurate forecasts are crucial for reducing sailing risks, protecting cargo, and supporting decarbonization efforts, as fuel consumption accounts for over 60\% of operational costs \cite{wang2018dynamic}.
Key weather parameters such as sea surface temperature (${\textrm{SST}}$), wave height, wave direction, wave period, and wind are critical for achieving these objectives for marine shipping \cite{chu2015fuel,zis2020ship}.
However, developing accurate forecasting models for various sectors remains challenging due to the inherent variability and uncertainty of weather systems \cite{Lorenz1963,lorenz1965study}.

Traditional weather forecasting relies on physics-based numerical weather prediction (NWP) models, which simulate atmospheric and oceanic conditions using supercomputers \cite{kalnay2003atmospheric,warner2010numerical}. Since the first successful numerical forecast in 1950 \cite{charney1950numerical}, NWP models have steadily improved through advances in numerical methods, physical parameterizations, and data assimilation \cite{bauer2015quiet}. Despite these improvements, NWP models often require post-processing to correct biases before they can be effectively applied across various sectors \cite{reikard2011forecasting,mathiesen2011evaluation,zhou2012forecast,theocharides2020day}. Traditionally, such corrections were made using statistical methods, but there has been a growing trend towards applying machine learning (ML) models \cite{o2018integrated,munoz2020aviation,alkhayat2021review,mayer2022benefits}.
As weather data accumulates over time, the expanding historical dataset enhances the accuracy of ML models. However, despite this growing abundance, the accumulated data cannot be directly used to improve the forecast accuracy of NWP models.
Additionally, the operational deployment of NWP models are often constrained by the high computational demands \cite{leutbecher2020probabilistic,benbouallegue2023rise}.

Recently ML models, developed with datasets like the European Centre for Medium-Range Weather Forecasts (ECMWF) ERA5 reanalysis data \cite{hersbach2020era5}, have shown significant promise in reducing computational demands and improving the timeliness and accuracy of forecasts \cite{de2023machine}. Models such as FourCastNet \cite{pathak2022fourcastnet}, Pangu-Weather \cite{bi2022panguweather}, GraphCast \cite{lam2022graphcast}, FuXi \cite{chen2023fuxi}, FengWu\cite{chen2023fengwu}, AIFS \cite{bouallegue2024aifs} and others are revolutionizing global weather forecasting \cite{ling2024artificial} by providing highly accurate forecasts with skills comparable to or even surpassing those of ECMWF's high-resolution forecasts (HRES) \citep{ECMWF2021}. Notably, these ML models are highly efficient, requiring only a single graphics processing unit (GPU) and less than a minute to generate a forecast. However, challenges remain in their practical applications.

One significant limitation of existing ML models is their 6-hour temporal resolution, which is insufficient for many applications requiring finer granularity. Although Pangu-Weather \cite{bi2022panguweather} introduces a hierarchical temporal aggregation strategy to enable 1 hourly forecasts, it struggles with prediction continuity, as generating forecasts for consecutive hours might require different numbers of iterations. For example, a 72-hour forecast requires 3 iterations, while a 71-hour forecast needs 8. This discontinuity in predictions compromises the models' reliability for weather forecasting and limits their applicability in scenarios requiring continuous predictions.
Therefore, there is an urgent need for methods that can enhance temporal resolution to 1 hour while ensuring prediction continuity. Video frame interpolation (VFI) \cite{dong2023video}, a technique from computer vision, aims to create intermediate frames, conceptually similar to temporal super-resolution in weather forecasting.
However, the efficacy of existing VFI techniques, whether those based on convolutional neural networks (CNNs) \cite{niklaus2017adaptive,niklaus2017separable,jiang2018super,bao2019depth}, which are criticized for their content-agnostic nature and inability to capture long-range dependencies, or the more recent VFI Transformer (VFIT) \cite{shi2022video,lu2022video}, limited to single-frame interpolation, remains insufficient to improve the temporal resolution of ML-based weather forecasting models.
This gap underscores the necessity for developing more advanced models capable of addressing the unique challenges of temporal super-resolution in weather forecasting. 
Additionally, current ML models often focus on a narrow set of meteorological variables, neglecting others that are crucial for applications with significant socio-economic impacts. For example, essential variables for wind and solar energy forecasting, such as 100-meter wind speed (${\textrm{WS100M}}$) and surface solar radiation downwards (${\textrm{SSRD}}$), as well as oceanic variables are not included in the ML outputs.
Recent efforts have aimed to address these gaps.
Wang et al. \cite{wang2024} coupled a ML atmosphere model and ML ocean model for seasonal climate forecasts.
Guo et al. \cite{guo2024} developed ORCA (Oceanic Reliable foreCAst) for global ocean circulation from multi-year to decadal time scales.
Chen et al. \cite{chen2024machine} introduced FuXi-S2S incorporating both atmospheric variables and ${\textrm{SST}}$ for subseasonal timescales. For medium-range scales, none of the current ML models include both atmospheric and ocean variable.

To address these limitations, we present FuXi-2.0, a significant advancement in ML-based weather forecasting. FuXi-2.0 is designed to generate global weather forecasts at a 1-hour temporal resolution for the first 5 days, transitioning to 6-hourly forecasts from day 5 to day 10. It consists of two models: one for generating 6-hourly forecasts and another for interpolating 1-hourly forecasts from the 6-hourly data. The model employs a transformer architecture for efficient interpolation, reducing the number of iterations and ensuring continuous 1-hourly forecasts. FuXi-2.0 also includes a broader range of variables, covering 5 upper-air atmospheric variables at 13 pressure levels and 23 surface variables, which are critical for applications in wind and solar energy, aviation, and marine shipping.
In particular, we include oceanic variables so that FuXi-2.0 can be regarded as the first atmosphere-ocean coupled model at the medium-range timescale.
Compared to Pangu-Weather, the only other ML-based model capable of generating 1-hourly forecasts, FuXi-2.0 demonstrates superior performance, outperforming both Pangu-Weather and ECMWF HRES in accuracy, underscoring its potential for various practical applications.
Particularly, using near surface wind forecasts from FuXi-2.0 for day-ahead wind power forecasting demonstrate better performance than using forecasts from ECMWF HRES.
Additionally, tropical cyclone (TC) forecasts are crucial for wind energy and marine shipping because they help in mitigating risks related to wind gusts, rapid wind direction changes, extreme waves, and heavy precipitation, which can severely damage offshore wind turbines \cite{hong2012economic,mattu2022impact,wen2024assessment,wang2024impact} and disrupt shipping operations \cite{zhang2024assessing}.
By incorporating ${\textrm{SST}}$ and 6 ocean surface wave variable as outputs in FuXi-2.0, we assess the feasibility of coupled ML prediction systems with ocean surface state information.
The FuXi-2.0 shows more accurate forecast in TC intensity than the FuXi-1.0 introduced by Chen et al. \cite{chen2023fuxi}, suggesting advantages of an atmosphere-ocean coupled model over atmosphere-only models.

\section{Results}

This study presents a thorough evaluation of the 1-hourly forecasts generated by FuXi-2.0 using testing data from 2018. The evaluation covers a broad range of variables, including those typically assessed in forecast evaluations and others crucial for wind and solar energy prediction, aviation, and marine shipping. The performance of FuXi-2.0 is compared with both Pangu-Weather and ECMWF HRES for variables commonly evaluated in ML-based weather forecasting models. For key variables essential for practical applications but not included in Pangu-Weather, FuXi-2.0 is compared exclusively with ECMWF HRES. This comprehensive assessment offers valuable insights into FuXi-2.0’s forecasting capabilities. ECMWF HRES provides 1-hourly forecasts for the first 90 hours, followed by 3-hourly forecasts up to 144 hours, and 6-hourly forecast from 144 to 240 hours. Consequently, our evaluation focuses on comparing the 1-hourly forecasts generated by FuXi-2.0 and ECMWF HRES for lead times ranging from 0 to 90 hours.

Typically, the ECMWF HRES analysis, denoted as HRES-fc0, serves as the benchmark when evaluating its performance. However, due to 4 daily initializations of ECMWF HRES (at 00/06/12/18 UTC), the temporal resolution of HRES-fc0 data is limited to 6 hours, making it unsuitable for evaluating 1-hourly forecasts. Therefore, ERA5 is used as the benchmark to assess the performance of ECMWF HRES, FuXi-2.0, and Pangu-Weather.
In the supplementary materials, where 6-hourly forecasts are discussed, HRES-fc0 data serves as the reference for evaluating the accuracy of ECMWF HRES, while ERA5 continues to be used for evaluating FuXi-2.0 and Pangu-Weather.

\subsection{Commonly evaluated variables}
In this subsection, we present a detailed evaluation of forecasting accuracy for variables commonly assessed in medium-range weather predictions, following the guidelines outlined in Weatherbench 2 \cite{rasp2023weatherbench}. Our analysis focuses on surface variables directly impacting human activities, including 2-meter temperature ($\textrm{T2M}$), mean sea level pressure ($\textrm{MSL}$), and 10-meter wind speed ($\textrm{WS10M}$), as well as upper-level variables crucial for understanding weather patterns and large-scale atmospheric evolution, including 850 hPa temperature ($\textrm{T850}$), 500 hPa geopotential ($\textrm{Z500}$), and 700 hPa specific humidity ($\textrm{Q700}$).
We exclude total precipitation ($\textrm{TP}$) forecasts due to known biases in ERA5 precipitation data \cite{Lavers2022}.

\begin{figure}[h]
    \centering
    \includegraphics[width=\linewidth]{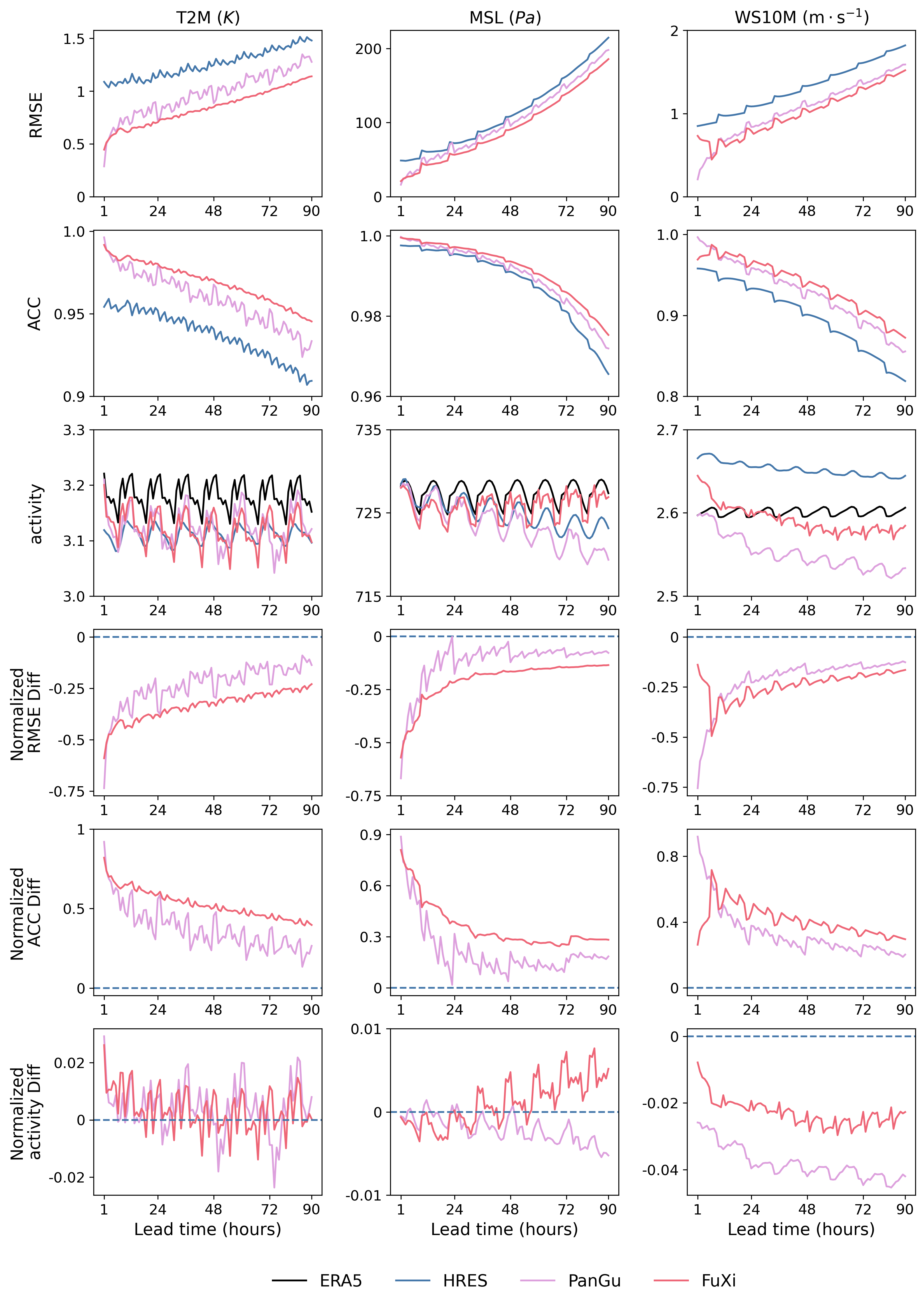}
    \caption{Comparison of globally-averaged and latitude-weighted root mean squared error ($\textrm{RMSE}$) (first row), anomaly correlation coefficient ($\textrm{ACC}$) (second row), and forecast/observation activity (third row), as well as normalized differences in $\textrm{RMSE}$ (fourth row), $\textrm{ACC}$ (fifth row), and activity (sixth row) of ECMWF HRES (blue lines), FuXi-2.0 (red lines), Pangu-Weather (purple lines), and ERA5 (black lines represent observation activity) for 3 surface variables: 2-meter temperature ($\textrm{T2M}$) (first column), mean sea level pressure ($\textrm{MSL}$) (second column), and 10-meter wind speed ($\textrm{WS10M}$) (third column), in 90-hour forecasts at a temporal resolution of 1 hour using testing data from 2018. The $\textrm{RMSE}$ and $\textrm{ACC}$ are calculated against ERA5, and normalized differences in $\textrm{RMSE}$, $\textrm{ACC}$, and activity are calculated using ECMWF HRES as the baseline model.}
    \label{common_variable_surface}    
\end{figure}

\begin{figure}[h]
    \centering
    \includegraphics[width=\linewidth]{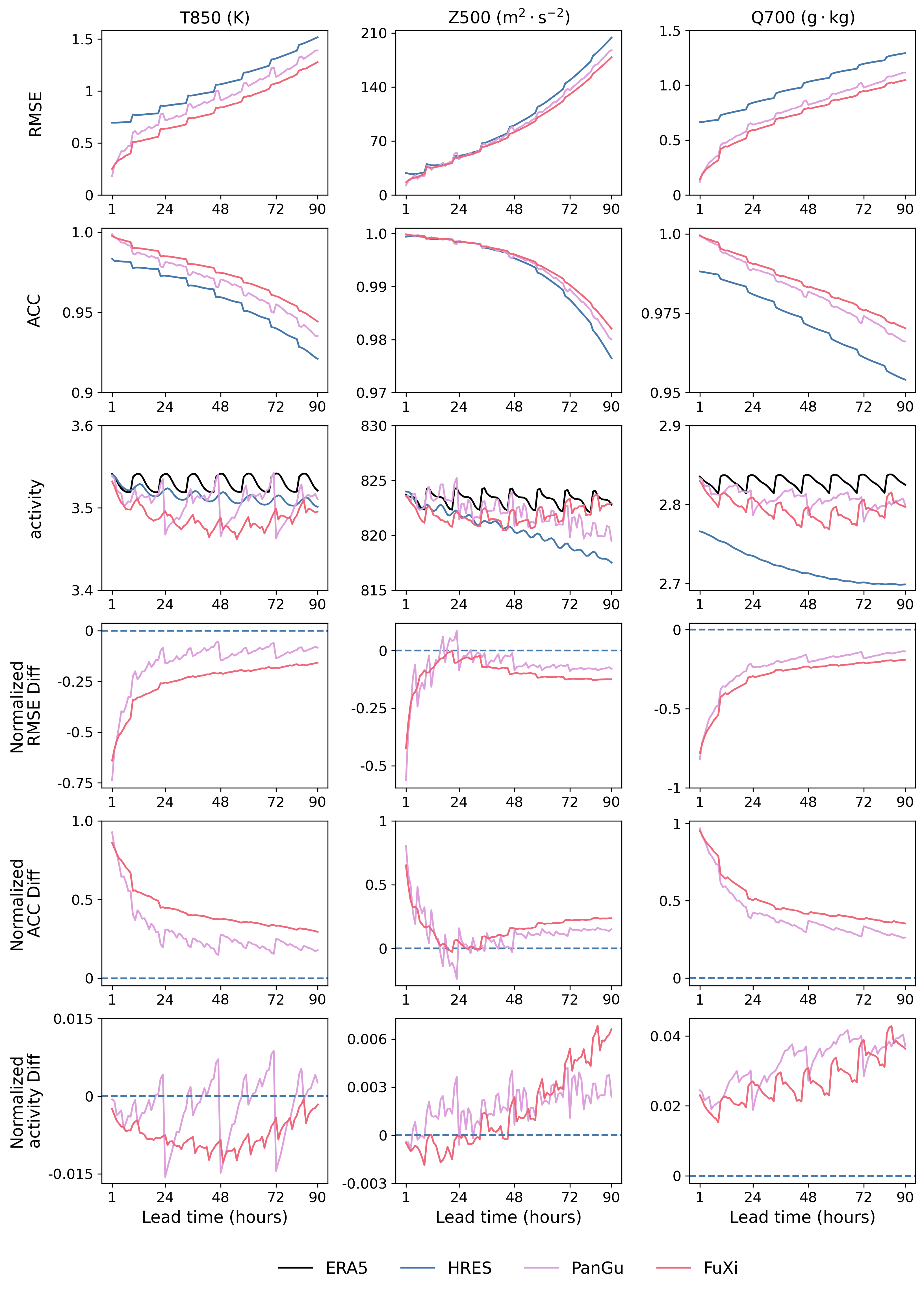}
    \caption{Comparison of globally-averaged and latitude-weighted root mean squared error ($\textrm{RMSE}$) (first row), anomaly correlation coefficient ($\textrm{ACC}$) (second row), and forecast/observation activity (third row), as well as normalized differences in $\textrm{RMSE}$ (fourth row), $\textrm{ACC}$ (fifth row), and activity (sixth row) of HRES (blue lines), FuXi-2.0 (red lines), and Pangu-Weather (purple lines), and ERA5 (black lines represent observation activity) for 3 upper-air variables, including 850 hPa temperature ${\textrm{T850}}$ (first column), 500 hPa geopotential ${\textrm{Z500}}$ (second column), 700 hPa specific humidity $\textrm{Q700}$ (third column), in 90-hour forecasts at a temporal resolution of 1 hour using testing data from 2018. The $\textrm{RMSE}$ and $\textrm{ACC}$ are calculated against ERA5, and normalized differences in $\textrm{RMSE}$, $\textrm{ACC}$, and activity are calculated using ECMWF HRES as the baseline model.}
    \label{common_variable_upper}    
\end{figure}

Figures \ref{common_variable_surface} and \ref{common_variable_upper} illustrate time series of globally-averaged and latitude-weighted root mean squared error ($\textrm{RMSE}$), anomaly correlation coefficient ($\textrm{ACC}$), and forecast activity for these variables.
Data from ECMWF HRES, FuXi-2.0, and Pangu-Weather, all with a temporal resolution of 1 hour, were analyzed.
Both FuXi-2.0 and Pangu-Weather outperform ECMWF HRES in $\textrm{RMSE}$ and $\textrm{ACC}$. 
Initially, Pangu-Weather shows a slight advantage over FuXi-2.0, but FuXi-2.0 performs better for most of the 90-hour forecast period, with lower $\textrm{RMSE}$ and higher $\textrm{ACC}$ across all examined variables and forecast lead times.
Notably, this trend of increasing advantage by FuXi-2.0 becomes more pronounced at longer forecast lead times. Additionally, the 4th and 5th rows of Figures \ref{common_variable_surface} and \ref{common_variable_upper} illustrate the normalized differences in $\textrm{RMSE}$ and $\textrm{ACC}$ compared to ECMWF HRES.
It is worth noting that forecasts produced by Pangu-Weather exhibit strong oscillations, due to the temporal aggregation strategy implemented by the Pangu-Weather model for generating 1-hourly forecasts.

While $\textrm{RMSE}$ and $\textrm{ACC}$ are critical metrics for assessing forecast quality, the interpretation of model's forecast performance should also consider other forecast characteristics, such as forecast activity, which is defined as standard deviation of the forecast anomaly (see subsection \ref{eval_method} for a formal definition).
Previous research has raised concerns that optimizing solely for $\textrm{RMSE}$ may lead to increasingly smooth forecast \cite{fuxi_extreme2023}. Therefore, our evaluation extends to forecast activity, comparing it against observed activity, as shown in the 3rd and 6th rows of Figures \ref{common_variable_surface} and \ref{common_variable_upper}.
The activity of ERA5 is also plotted in the Figures as a reference.
We find that the ERA5 generally has the highest activity except for the $\textrm{WS10}$ for which ECMWF HRES has the highest activity values likely due to its original 0.1\textdegree spatial resolution. 
Overall, the Pangu-Weather has the lowest activity for surface variables, suggesting the forecasts from Pangu-Weather are smoother than those from ECMWF HRES and FuXi, which is particularly pronounced over longer lead times.
For upper-air variables, the temporal variation of forecast activity of FuXi-2.0 is closer to that of ERA5.
Pangu-Weather shows stronger variations, and the overall activity of FuXi-2.0 and Pangu-Weather is similar and higher than that of ECMWF HRES occasionally.
However, for many applications, overly smooth forecast are not desired.
Notably, FuXi-2.0 exhibits higher activity than Pangu-Weather while still achieving better performance in $\textrm{RMSE}$ and $\textrm{ACC}$, demonstrating that a better $\textrm{RMSE}$ does not necessarily come at the cost of unrealistically low forecast activity due to smoothing.
It is important to note that ECMWF HRES serves as the baseline model for computing normalized RMSE, ACC, and activity differences.
Detailed results for 6-hourly forecasts are available in the supplementary materials (See Supplementary Figures 1 and 2). 

\FloatBarrier

\FloatBarrier

Forecast smoothness is also assessed through power spectra, which reveal energy distribution across different scales and illustrate differences in forecast detail and variability.
Higher power at smaller scales suggests finer-scale structures. Following Rasp et al. \cite{rasp2023weatherbench}, we calculated mean zonal energy spectra within the latitudinal bands from 30\textdegree N (S) to 60\textdegree N (S). 
Supplementary Figure 3 displays power spectra for 6 variables across 5 forecast lead times (12 24, 48, 72, and 90 hours).
ERA5 spectra remain consistent, serving as a baseline for assessing forecasts smoothness over time.
ECMWF HRES shows the highest power spectra for 5 variables except $\textrm{Z500}$ throughout the forecast lead times, particularly at shorter wavelengths, demonstrating the strengths of physics-based models.
For $\textrm{Z500}$, FuXi-2.0 shows higher energy at smaller wavelengths than ERA5, Pangu-Weather, and ECMWF HRES.
Both FuXi-2.0 and Pangu-Weather closely align with ERA5, though their performance at smaller scales degrades with longer lead times, with Pangu-Weather spectra shows higher degree of degradation, particularly in $\textrm{Q700}$ forecasts.
This decline across all scales underscores the challenges in accurately capturing the fine-scale details.
Moreover, both FuXi-2.0 and Pangu-Weather yield quite similar spectra with the FuXi-2.0 model showing slightly higher energy than Pangu-Weather at all scales, exhibiting a slightly better agreement with ERA5 than the Pangu-Weather model.
The superiority of FuXi-2.0 over Pangu-Weather becomes more pronounced with increasing forecast lead times, indicating its greater efficacy in improving forecasts for small-scale weather patterns.
 
\subsection{Wind and solar energy, aviation, and shipping forecast performance}

This subsection analyzes the forecasting accuracy of variables critical for wind and solar energy, aviation, and marine shipping.
For wind energy, we focus on the 100-meter u wind component (${\textrm{U100M}}$), 100-meter v wind component (${\textrm{V100M}}$), and 100-meter wind speed (${\textrm{WS100M}}$).
In solar energy, our analysis targets surface net solar radiation (${\textrm{SSR}}$), surface solar radiation downwards (${\textrm{SSRD}}$), and total sky direct solar radiation at surface (${\textrm{FDIR}}$).
Aviation analysis focuses on cloud cover at various levels: low cloud cover (${\textrm{LCC}}$), medium cloud cover (${\textrm{MCC}}$), high cloud cover (${\textrm{HCC}}$), and total cloud cover (${\textrm{TCC}}$). 
Visibility, while crucial for aviation, is excluded from FuXi-2.0 due to its absence in the ERA5 reanalysis dataset.
Marine shipping analyses incorporate essential oceanic parameters such as sea surface temperature (${\textrm{SST}}$), mean direction of total swell (${\textrm{MDTS}}$), mean direction of wind waves (${\textrm{MDWW}}$), mean period of total swell (${\textrm{MPTS}}$), mean period of wind waves (${\textrm{MPWW}}$), significant height of total swell (${\textrm{SHTS}}$), and significant height of wind waves (${\textrm{SHWW}}$).
These variables are assessed to determine the reliability and enhancement of shipping safety and efficiency offered by FuXi-2.0

\begin{figure}[h]
    \centering
    \includegraphics[width=\linewidth]{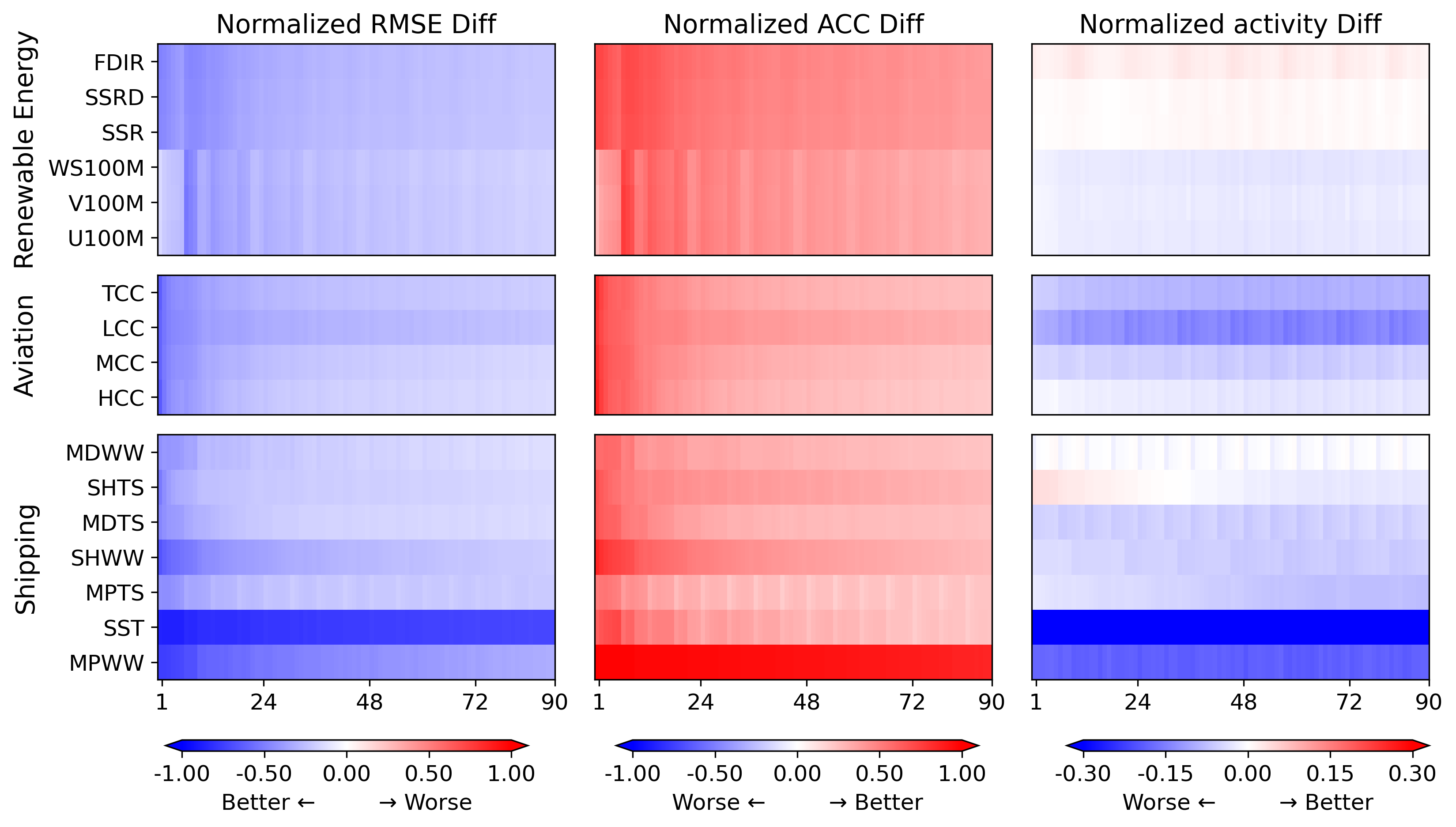}
    \caption{Comparison of normalized differences in globally-averaged and latitude-weighted root mean squared error $\textrm{RMSE}$ (first column), anomaly correlation coefficient $\textrm{ACC}$ (second column), and forecast activity (third column) of FuXi-2.0 compared to ECMWF HRES, in 90-hour forecasts at a temporal resolution of 1 hour using testing data from 2018. The evaluation encompasses a broad range of variables relevant to different sectors. For wind and solar energy forecasting (first row), variables include 100-meter u wind component (${\textrm{U100M}}$), 100-meter v wind component (${\textrm{V100M}}$), 100-meter wind speed (${\textrm{WS100M}}$), surface net solar radiation (${\textrm{SSR}}$), surface solar radiation downwards (${\textrm{SSRD}}$), and total sky direct solar radiation at surface (${\textrm{FDIR}}$). In the aviation sector (second row), variables include low cloud cover (${\textrm{LCC}}$), medium cloud cover (${\textrm{MCC}}$), high cloud cover (${\textrm{HCC}}$), and total cloud cover (${\textrm{TCC}}$). For marine shipping (third row), variables evaluated include sea surface temperature (${\textrm{SST}}$), mean direction of total swell (${\textrm{MDTS}}$), mean direction of wind waves (${\textrm{MDWW}}$), mean period of total swell (${\textrm{MPTS}}$), mean period of wind waves (${\textrm{MPWW}}$), significant height of total swell (${\textrm{SHTS}}$), and significant height of wind waves (${\textrm{SHWW}}$). All variables are evaluated at a temporal resolution of 1 hour using testing data from 2018. The $\textrm{RMSE}$ and $\textrm{ACC}$ are calculated against ERA5, and normalized differences in $\textrm{RMSE}$, $\textrm{ACC}$, and activity are calculated using ECMWF HRES as the baseline model.}
    \label{normalized_difference}    
\end{figure}

Figure \ref{normalized_difference} presents the normalized differences in globally-averaged and latitude-weighted $\textrm{RMSE}$, $\textrm{ACC}$, and forecast activity for 1-hourly forecasts, comparing the performance of FuXi-2.0 with ECMWF HRES using one-year testing data from 2018.
ECMWF HRES serves as the baseline for calculating these normalized differences.
Lower $\textrm{RMSE}$ values indicate superior forecast performance, with negative values of $\textrm{RMSE}$ differences suggesting that FuXi-2.0 produces more accurate forecasts than ECMWF HRES, while positive values indicate the opposite.
In contrast, higher $\textrm{ACC}$ values reflect better alignment of forecast anomalies with observed anomalies, with positive $\textrm{ACC}$ differences favoring FuXi-2.0's performance.
Furthermore, higher forecast activity values represent an enhanced capability in capturing significant weather variations, with positive differences signifying greater variability in FuXi-2.0’s forecasts compared to ECMWF HRES.
Over forecast lead times from 0 to 90 hours, FuXi-2.0 consistently outperforms ECMWF HRES in all examined weather parameters, as evidenced by negative $\textrm{RMSE}$ differences (in blue) and positive $\textrm{ACC}$ differences (in red).
Regarding forecast activity, FuXi-2.0 demonstrates increased variability in solar radiation forecasts relative to ECMWF HRES, whereas it shows less variability in wind and cloud cover parameters compared to ECMWF HRES.
In marine shipping, FuXi-2.0 provides notably smoother forecasts for $\textrm{SST}$ and $\textrm{MPWW}$, with negligible differences observed in other oceanic variables.
Detailed results for 6-hourly forecasts of the same variables are shown in Supplementary Figure 4.

While FuXi-2.0 provides more accurate forecasts for wind speed and solar radiation, we now determine whether these forecasts provide practical benefits over ECMWF HRES in reducing operational costs and boosting wind energy integration into the power grid.


\FloatBarrier

\subsection{Wind power forecast performance}

\begin{figure}[h]
    \centering
    \includegraphics[width=\linewidth]{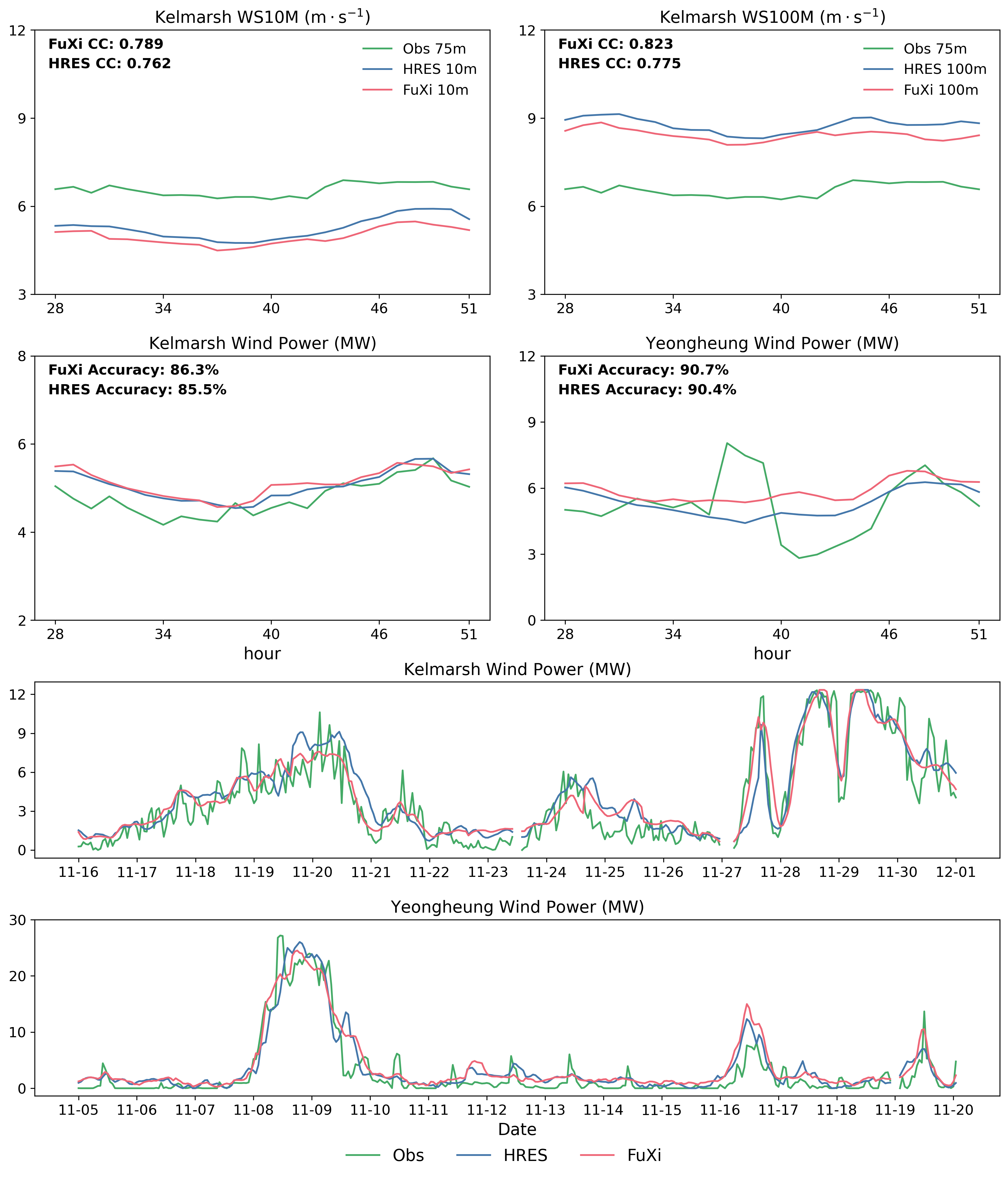}
    \caption{Comparison of wind speed and wind power at Kelmarsh and Yeongheung wind farms. The top row illustrates one-year averaged 10-meter wind speed ($\textrm{WS10M}$) and 100-meter wind speed ($\textrm{WS100M}$) as a function of forecast lead times for Kelmarsh wind farm. The second rows shows one-year averaged wind power forecasts as a function of forecast lead times, for both Kelmarsh and Yeongheung wind farms. The third and fourth row present example two 15-day time series of observed and predicted wind power from November 16th to December 1st, 2018, for Kelmarsh, and from November 5th to November 20th, 2018, for Yeongheung, respectively. The forecast horizon for these day-ahead wind power forecasts ranges from 28 to 51 hours.}
    \label{wind_power}    
\end{figure}

Figure \ref{wind_power} compares wind speed and wind power at the Kelmarsh and Yeongheung wind farms.
Wind speed data is available only for Kelmarsh (at a turbine height of 75 meters), limiting the comparison of wind speeds to this farm.
FuXi wind speed forecasts at 10 meters ($\textrm{WS10M}$) and 100 meters ($\textrm{WS100M}$), show higher correlation coefficients (CC) than ECMWF HRES forecasts.
The figure also illustrates the performance of ML-based day-ahead wind power forecasting models using data from both FuXi-2.0 and ECMWF HRES, with forecast horizons ranging from 28 to 51 hours.
Both models have identical structures and training procedures (see subsections \ref{power_model} and \ref{eval_method}).
Results demonstrate that using FuXi-2.0 as inputs lead to more accurate wind power forecasts at both farms, outperforming those based on ECMWF HRES.
This highlights FuXi-2.0 as a promising alternative, given ECMWF HRES is widely used by many power forecasting service companies.
Additionally, Figure \ref{wind_power} includes two 15-day time series showing observed and predicted wind power.
The model using FuXi-2.0 forecasts, unlike HRES, successfully captures the timing of a significant wind power down-ramp event at Kelmarsh between November 20th and 21st.

\FloatBarrier

\subsection{Tropical cyclone forecast performance}

TCs are among the most destructive extreme events, posing severe risks to human lives and property \cite{mousavi2011global,lang2012impact}.
Enhancing the accuracy of TC track forecasts can save lives or reduce unnecessary warnings and evacuations, thereby improving emergency preparedness and management \cite{hamill2011global,conroy2023track}.
These forecasts are also crucial for offshore wind farms, allowing for timely shutdowns to prevent turbine damage and ensure crew safety. In marine shipping, accurate forecasts enable route adjustments to avoid hazardous conditions, protecting both cargo and crew.

TCs derive their energy primarily from ocean heat. Forecasts that do not couple atmospheric and oceanic processes overlook crucial feedback from surface heat exchange.
Previous studies using traditional NWP models have demonstrated that atmosphere-ocean coupling improves TC intensity forecasts \cite{mogensen2017tropical}.
This coupling is already operational in ECMWF’s ensemble forecasts and is planned for inclusion in the HRES in future updates.
By integrating ${\textrm{SST}}$ and ocean surface wave variables into FuXi-2.0, we explore the potential of coupled ML models to enhance TC prediction accuracy.

\begin{figure}
    \centering
    \includegraphics[width=\linewidth]{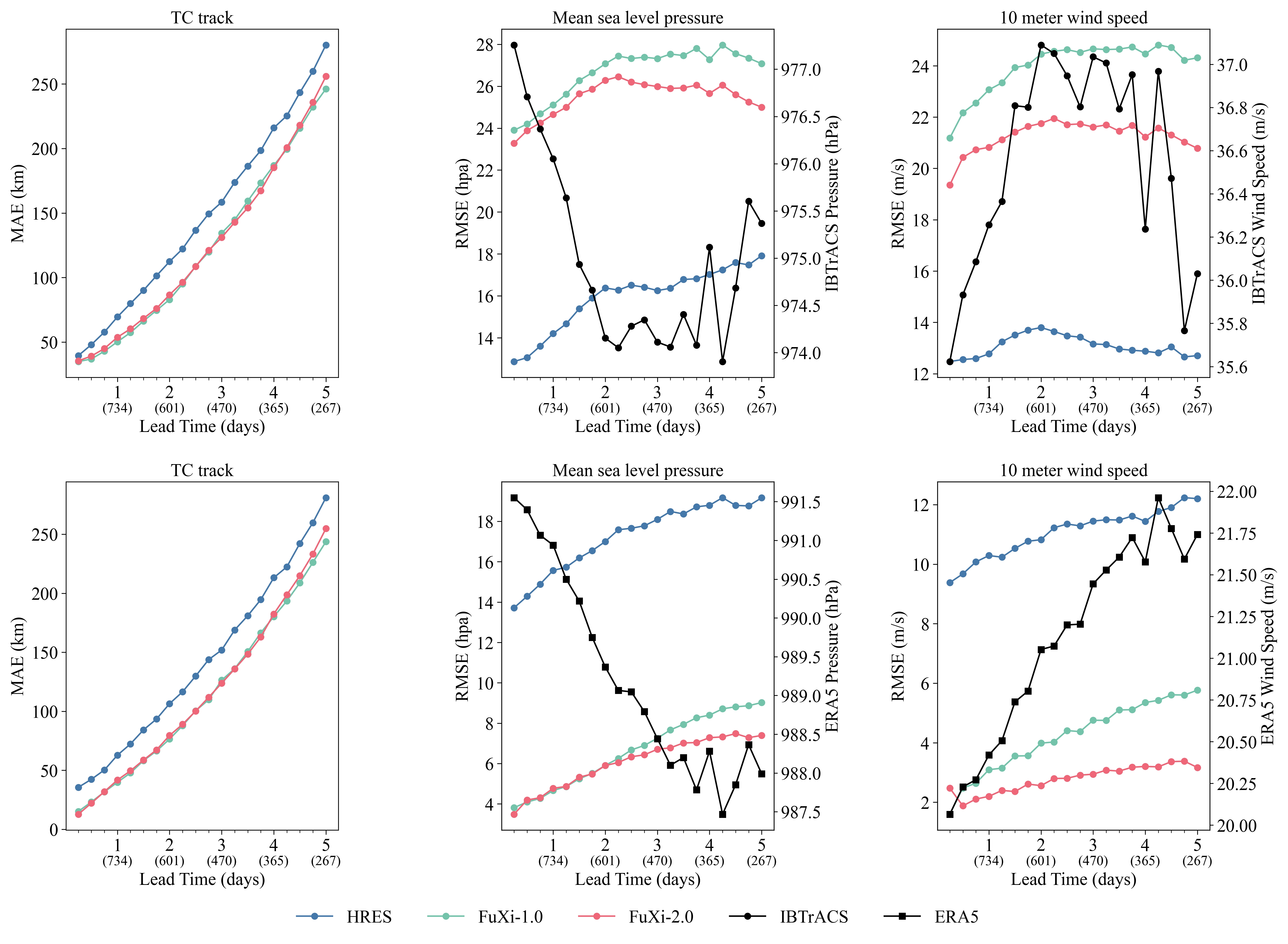}
    \caption{Comparison of the average mean absolute error (MAE) for tropical cyclone (TC) track forecasts (first column) and root mean squared error (RMSE) for mean sea-level pressure (${\textrm{MSL}}$) (second column) and 10-meter wind speed ($\textrm{WS10M}$) (third column) in TC intensity forecasts for HRES (blue lines), FuXi-1.0 (green lines), FuXi-2.0 (red lines), Pangu-Weather (purple lines), International Best Track Archive for Climate Stewardship (IBTrACS) (black lines with solid circles), and ERA5 (black lines with squares), as a function of forecast lead times. The evaluation covers forecasts for 90 TCs, and is performed against the IBTrACS dataset. The $\textrm{MSL}$ and $\textrm{WS10M}$ forecast comparisons are dual Y axis figures, here the secondary Y axis shows the IBTrACS data (black lines with solid circles) in the first row and ERA5 (black lines with squares) in the second row. Numbers in brackets on the x-axis indicate the sample size for each forecast lead time. For example, ‘(734)’ means 734 samples from FuXi-1.0, FuXi-2.0, and ECMWF-HRES were averaged from 734 initial points where the TC persisted for at least 24 hours.}
    \label{TC_stat}
\end{figure}

This study compares the forecast performance of FuXi-2.0, which couples atmospheric and oceanic variables, with that of FuXi-1.0 which models atmosphere alone.
The comparison focuses on track and intensity predictions for 90 TCs in 2018.
Figure \ref{TC_stat} presents a statistical analysis of 5-day forecasts generated by HRES, FuXi-1.0, and FuXi-2.0, evaluated against two benchmarks: the International Best Track Archive for Climate Stewardship (IBTrACS) dataset (top row) and TC data extracted from the ERA5 dataset (second row).
For track forecasts, FuXi-1.0 and FuXi-2.0 demonstrate similar performance, both slightly surpassing HRES, consistent with findings of Zhong et al. \cite{zhong2023fuxiextreme}.
In terms of intensity predictions, FuXi-2.0 shows improvements over FuXi-1.0, particularly in maximum $\textrm{WS10M}$ and minimum $\textrm{MSL}$, as evidenced by reduced RMSE values, regardless if the forecasts are evaluated against the IBTrACS or ERA5 datasets.
The performance enhancement of FuXi-2.0 relative to FuXi-1.0 increases with longer forecast lead times.
ECMWF HRES consistently exhibits superior performance in TC intensity forecasts when evaluated against the IBTrACS dataset, achieving the lowest RMSE values throughout the forecast period.
However, when referenced against the ERA5 dataset, both FuXi-1.0 and FuXi-2.0 outperform ECMWF HRES, with FuXi-2.0 consistently providing the most accurate intensity forecasts.
This discrepancy likely arises from differences in TC intensity between the ERA5 and IBTrACS datasets, where ERA5 generally records weaker TC intensities, characterized by higher $\textrm{MSL}$ and lower $\textrm{WS10M}$ values.
These findings align with the forecast activity comparisons shown in Figure \ref{common_variable_surface}, where ECMWF HRES shows significantly higher activity compared to ERA5 and FuXi-2.0 for $\textrm{WS10M}$.
In summary, our results demonstrate the coupled atmosphere-ocean model employed by FuXi-2.0 have advantages in TC intensity forecast over the atmosphere-only model used in FuXi-1.0.

\section{Discussion} 

Recent advances in ML for weather forecasting models have led to significant achievements, with state-of-the-art ML models surpassing the ECMWF HRES in global weather prediction accuracy \cite{bi2022panguweather,lam2022graphcast,chen2023fuxi,olivetti2023advances}.
These ML models typically offer forecasts with a temporal resolution of 6 hours and a spatial resolution of $0.25^{\circ}$.
However, a notable limitation is their 6-hour temporal resolution and their restricted focus on a limited set of basic meteorological variables, neglecting many variables that are important for a wide range of applications.
This raises important questions about the future role of ML models in weather forecasting, particularly regarding their ability to provide reliable and valuable contributions to weather services and reshape the forecasting landscape.

In this paper, we introduce FuXi-2.0, an advancement in ML-based weather forecasting model designed to enhance the practicality and accuracy of forecasts. FuXi-2.0 introduces an innovative framework composed of two models: one generating forecasts at 6-hour intervals and another interpolating these forecasts to produce continuous 1-hourly predictions.
This approach ensures consistent 1-hourly forecasts while expanding the range of included variables to address the needs of sectors such as wind and solar energy, aviation, and maritime shipping.
Our results demonstrate that FuXi-2.0 consistently outperforms ECMWF HRES in forecasting key meteorological variables relevant to these sectors.
Notably, FuXi-2.0 exhibits superior performance in wind power forecasting, further validating its effectiveness as a reliable tool for scenarios requiring accurate weather predictions.

Additionally, research indicates that anthropogenic climate warming and rising SSTs are likely to intensify TCs and increase their associated rainfall \cite{pielke2005hurricanes,kossin2013trend,patricola2018anthropogenic,Knutson2019}.
Consequently, accurate TC forecasts are increasingly critical for ensuring timely and effective warnings.
FuXi-2.0 integrates both atmospheric and oceanic data, laying the foundation for the development of comprehensive atmospheric-oceanic coupled models. This integration represents a significant advancement in the field, with FuXi-2.0 delivering improved TC intensity forecasts compared to its predecessor, FuXi-1.0, through the inclusion of ocean surface variables.

Future enhancements to FuXi model are planned to further improve its accuracy and utility. These enhancements include:

\begin{itemize}

\item Expanding the application of FuXi-2.0 forecasts to additional industries, similar to its current use in wind power forecasting, to demonstrate the model’s effectiveness in practical applications.

\item Increasing spatial resolution from $0.25^{\circ}$ to higher resolutions and extending the model's top levels to higher altitudes, reaching beyond the stratosphere and into the mesosphere.

\item Enhancing the model’s utility for practical applications by incorporating a broader range of variables. While certain diagnostic variables are critical for downstream tasks, they will not be directly included in the model's output. Instead, we will develop task-specific models that utilize FuXi-2.0's output to derive these variables.

\item Transitioning from the ERA5 dataset to more accurate data sources, such as surface station measurements, to improve precipitation accuracy. Current global precipitation datasets, such as those from GPM \cite{huffman2015nasa} and TRMM, often underrepresent precipitation intensity. To address this, we plan to include rainfall categorization in the model outputs using quantile thresholds derived from historical data analysis.

\end{itemize}

Our goal is to establish partnerships with industry professionals interested in exploring the diverse applications of the FuXi model, leveraging its enhanced forecasting capabilities for real-world scenarios.

\section{Methods} 
 
\subsection{Data}
The ERA5 reanalysis dataset \cite{hersbach2020era5}, produced by the European Centre for Medium-Range Weather Forecasts (ECMWF), provides hourly data from January 1950 onwards, with a spatial resolution of approximately 31 km. This dataset, recognized for its extensive coverage and high accuracy, forms the basis of our study. We utilized two ERA5 subsets with a 0.25\textdegree spatial resolution, corresponding to $721\times1440$ grid points. The first subset covers a 8-year period from 2012 to 2017 for training the 6-hourly model, while the second, spanning from 2015 to 2017, is used for train the 1-hourly model. FuXi-2.0's performance was evaluated using a separate 1-year dataset from 2018. 

FuXi-2.0 forecasts 88 meteorological variables, including 5 upper-air atmospheric variables across 13 pressure levels (50, 100, 150, 200, 250, 300, 400, 500, 600, 700, 850, 925, and 1000 hPa), and 23 surface variables. Upper-air variables include geopotential (${\textrm{Z}}$), temperature (${\textrm{T}}$), u component of wind (${\textrm{U}}$), v component of wind (${\textrm{V}}$), and specific humidity (${\textrm{Q}}$). Surface variables comparise 2-meter temperature (${\textrm{T2M}}$), 2-meter dewpoint temperature (${\textrm{D2M}}$), sea surface temperature (${\textrm{SST}}$), 10-meter u wind component (${\textrm{U10M}}$), 10-meter v wind component (${\textrm{V10M}}$), 100-meter u wind component (${\textrm{U100M}}$), 100-meter v wind component (${\textrm{V100M}}$), mean sea-level pressure (${\textrm{MSL}}$), low cloud cover (${\textrm{LCC}}$), medium cloud cover (${\textrm{MCC}}$), high cloud cover (${\textrm{HCC}}$), total cloud cover (${\textrm{TCC}}$), surface net solar radiation (${\textrm{SSR}}$), surface solar radiation downwards (${\textrm{SSRD}}$), total sky direct solar radiation at surface (${\textrm{FDIR}}$), top net thermal radiation (${\textrm{TTR}}$) \footnote{${\textrm{TTR}}$ is also known as the negative of outgoing longwave radiation (${\textrm{OLR}}$).}, total precipitation ${\textrm{TP}}$, mean direction of total swell (${\textrm{MDTS}}$), mean direction of wind waves (${\textrm{MDWW}}$), mean period of total swell (${\textrm{MPTS}}$), mean period of wind waves (${\textrm{MPWW}}$), significant height of total swell (${\textrm{SHTS}}$), and significant height of wind waves (${\textrm{SHWW}}$). A comprehensive list of these variables and their abbreviations is detailed in Table \ref{glossary}. 

In addition to time-varying meteorological data that serve dual roles as input and output, the FuXi-2.0 model integrates static and temporal encodings of geographical and temporal information, such as latitude, longitude, daily and seasonal variations, orography, land-sea mask, and forecasting steps. These encoded features are utilized exclusively as input data to enhance the model's predictive capability.

For operational use, FuXi-2.0 primarily utilizes analysis data to avoid the approximately 5-day delay characteristic of ERA5 reanalysis data.
Accumulated variables including ${\textrm{TP}}$, ${\textrm{SSR}}$, ${\textrm{SSRD}}$, ${\textrm{FDIR}}$, and ${\textrm{TTR}}$ in the ERA5 reanalysis data, are accumulated over a specific time interval and possibly have non-zero values.
In contrast, in the analysis data, such accumulated variables are initialized to zero, since accumulation has not started yet at model initialization time.
As a result, to ensure consistency with operational implementations, all the accumulated variables are reset to zero before the model training.
Additionally, the 7 oceanic variables (${\textrm{SST}}$, ${\textrm{MDTS}}$, ${\textrm{MDWW}}$, ${\textrm{MPTS}}$, ${\textrm{MPWW}}$, ${\textrm{SHTS}}$, and ${\textrm{SHWW}}$), which are undefined over land, are systematically set to not-a-number (NaN) to maintain data integrity and consistency.

\begin{table}
\centering
\caption{\label{glossary} A summary of all the input and output variables. The "Type" indicates whether the variable is a time-varying variable including upper-air, surface, and geographical variables, or a temporal variable. The “Full name” and “Abbreviation” columns refer to the complete name of each variables and their corresponding abbreviations in this paper. The "Role" column clarifies whether each variable serves as both an input and a output, or is solely utilized as an input by our model.}
\begin{tabularx}{\textwidth}{cXccc}
\hline
\textbf{Type} & \textbf{Full name} & \textbf{Abbreviation} & \textbf{Role} \\
\hline
upper-air           & geopotential & ${\textrm{Z}}$ & Input and Output \\
                    & temperature  & ${\textrm{T}}$ & Input and Output \\
                    & u component of wind & ${\textrm{U}}$ & Input and Output \\
                    & v component of wind & ${\textrm{V}}$ & Input and Output \\
                    & specific humidity & ${\textrm{Q}}$ & Input and Output \\
                    \hline
surface             & 2-meter temperature & ${\textrm{T2M}}$ & Input and Output \\
                    & 2-meter dewpoint temperature & ${\textrm{D2M}}$ & Input and Output \\
                    & sea surface temperature & ${\textrm{SST}}$ & Input and Output \\
                    & 10-meter u wind component & ${\textrm{U10M}}$ & Input and Output \\
                    & 10-meter v wind component & ${\textrm{V10M}}$ & Input and Output \\
                    & 100-meter u wind component & ${\textrm{U100M}}$ & Input and Output \\
                    & 100-meter v wind component & ${\textrm{V100M}}$ & Input and Output \\
                    & mean sea-level pressure & ${\textrm{MSL}}$ & Input and Output \\
                    & low cloud cover & ${\textrm{LCC}}$ & Input and Output \\           
                    & medium cloud cover & ${\textrm{MCC}}$ & Input and Output \\             
                    & high cloud cover & ${\textrm{HCC}}$ & Input and Output \\            
                    & total cloud cover & ${\textrm{TCC}}$ & Input and Output \\
                    & surface net solar radiation & ${\textrm{SSR}}$ & Input and Output \\
                    & surface solar radiation downwards & ${\textrm{SSRD}}$ & Input and Output \\
                    & total sky direct solar radiation at surface & ${\textrm{FDIR}}$ & Input and Output \\
                    & top net thermal radiation & ${\textrm{TTR}}$ & Input and Output \\            
                    & total precipitation & ${\textrm{TP}}$ & Input and Output \\    
\hline                    
wave                & mean direction of total swell & ${\textrm{MDTS}}$ & Input and Output \\
                    & mean direction of wind waves & ${\textrm{MDWW}}$ & Input and Output \\   
                    & mean period of total swell & ${\textrm{MPTS}}$ & Input and Output \\
                    & mean period of wind waves  & ${\textrm{MPWW}}$ & Input and Output \\ 
                    & significant height of total swell & ${\textrm{SHTS}}$ & Input and Output \\
                    & significant height of wind waves & ${\textrm{SHWW}}$ & Input and Output \\
\hline
geographical        & orography & ${\textrm{OR}}$ & Input \\
                    & land-sea mask & ${\textrm{LSM}}$ & Input \\
                    & latitude & ${\textrm{LAT}}$ & Input \\
                    & longitude & ${\textrm{LON}}$ & Input \\  
\hline                  
temporal            & hour of day & ${\textrm{HOUR}}$ & Input \\
                    & day of year & ${\textrm{DOY}}$ & Input \\
                    & step & ${\textrm{STEP}}$ & Input \\                   
\hline
\end{tabularx}
\end{table}

In the wind energy sector, accurate wind power forecasting is more critical for effective grid management than mere wind speed predictions.
This study utilizes data from two wind farms: Kelmarsh wind farm in the United Kingdom, provided by Cubico Sustainable Investments \cite{Kelmarsh2022}, and Yeongheung wind farm in South Korea \cite{Yeongheung2018,JeongRim2022}, located near the the FuXi-2.0 model grid points at 52.5\textdegree N, 1\textdegree W (Kelmarsh) and 37.25\textdegree N, 126.5\textdegree E (Yeongheung), respectively.
The Kelmarsh dataset, spanning from 2016 to mid 2021, includes 10-minute interval measurements from six 2.05 megawatts (MW) wind turbines.
The Yeongheung dataset includes 1-hour interval data from two wind power complexes: Complex 1 with 9 wind turbines (22 MW total) and Complex 2 with 8 wind turbines (24 MW total).
For the development of our wind power forecasting model, data from January 1 to October 31, 2018, were used for training, while data from November 1 to December 31, 2018, were reserved for testing.
Expanding the study to additional wind farms is constrained by data availability and turbine coordinates.

To ensure data integrity, the following quality control steps were taken, reducing the Kelmarsh dataset from 8,760 to 8,319 and the Yeongheung dataset from 8760 to 8476 samples (see Supplementary Figure 5):
\begin{enumerate}
\item Established a 90th percentile wind speed threshold for zero wind power output.
\item Removed data where power was zero but wind speed exceeded this threshold.
\item Removed data where power was non-zero but wind speed was below this threshold.
\item Excluded data where both power and wind speed were zero.
\end{enumerate}
Since Yeongheung wind speed observations are not available from open sources, ECMWF HRES data are used as wind speed observations for data quality control.

Although ERA5 data is typically used for training and evaluating ML-based weather forecasting models, they may not always be the most accurate data source for certain applications.
Surface weather stations provide more accurate data, but are sparsely and irregularly distributed.
Jiang et al. \cite{jiang2021} demonstrated that the High-Resolution China Meteorological Administration (CMA) Land Data Assimilation System (HRCLDAS) \cite{han2019development} offer lower bias and higher correlation with observed 10-meter wind data compared to ERA5.
The HRCLDAS dataset integrates ground station observations, numerical weather prediction (NWP) data, and satellite data through a multi-grid variational analysis technique \cite{xie2011space} and a terrain correction algorithm \citep{shi2017statistical}.
Despite its regional limitation to East Asia, HRCLDAS provides a valuable benchmark.
In this study, we analyzed 1-hourly ${\textrm{WS10M}}$ data from ECMWF's high-resolution forecasts (HRES) \cite{ECMWF2021}, initialized twice daily at 00 and 12 UTC, with lead times from 1 to 12 hours.
Although ECMWF HRES is also initialized at 06 and 18 UTC, historical data for these initialization times are unavailable.
The spatial correlation and normalized differences between ERA5 and ECMWF HRES against HRCLDAS data are illustrated in Supplementary Figure 6.
The color coding in the normalized differences are blue, red and white, indicating areas where ECMWF HRES has higher, lower, or the same correlation coefficient with HRCLDAS data compared to ERA5.
Predominantly blue areas suggest ECMWF HRES has superior accuracy than ERA5 for $\textrm{WS10M}$ over most of East Asia.
Therefore, for FuXi-2.0 development, we used ${\textrm{U100M}}$, ${\textrm{V100M}}$, ${\textrm{U10M}}$, and ${\textrm{V10M}}$ data from ECMWF HRES forecasts with 1-12 hour lead times initialized at 00 and 12 UTC.

For evaluating tropical cyclone (TC) track forecasts, we used the International Best Track Archive for Climate Stewardship (IBTrACS) \cite{knapp2010,knapp2018} dataset, which provides global TC tracks at 6-hour intervals.
A modified ECMWF TC tracker, as developed by Zhong et al. \cite{zhong2023fuxiextreme}, was employed to extract and assess TC tracks from the ECMWF HRES, FuXi-1.0 \cite{chen2023fuxi}, FuXi-2.0, and ERA5 datasets.


\subsection{FuXi-2.0 model}

The FuXi-2.0 model advances ML-based weather forecasting by addressing the limitations of previous models which primarily only provided 6-hourly forecasts, which are insufficient for many applications.
Bi et al. \cite{bi2022panguweather} introduced the Pangu-Weather model, which employs a hierarchical temporal aggregation strategy using four distinct models for 1-hour, 3-hour, 6-hour, and 24-hour forecasts, respectively.
While this strategy enables hourly forecasts, it encounters challenges in prediction continuity due to the varying number of iterations required for consecutive hourly forecasts.
To overcome these challenges, the FuXi-2.0 model integrates a novel dual-model system that delivers continuous 1-hourly forecasts.
This system consists of a primary model optimized for 6-hourly forecasts and a secondary model that interpolates hourly forecasts from these 6-hourly forecasts.
This architecture significantly enhances reliability and expands the applicability across various scenarios.

\begin{figure}[h]
    \centering
    \includegraphics[width=\linewidth]{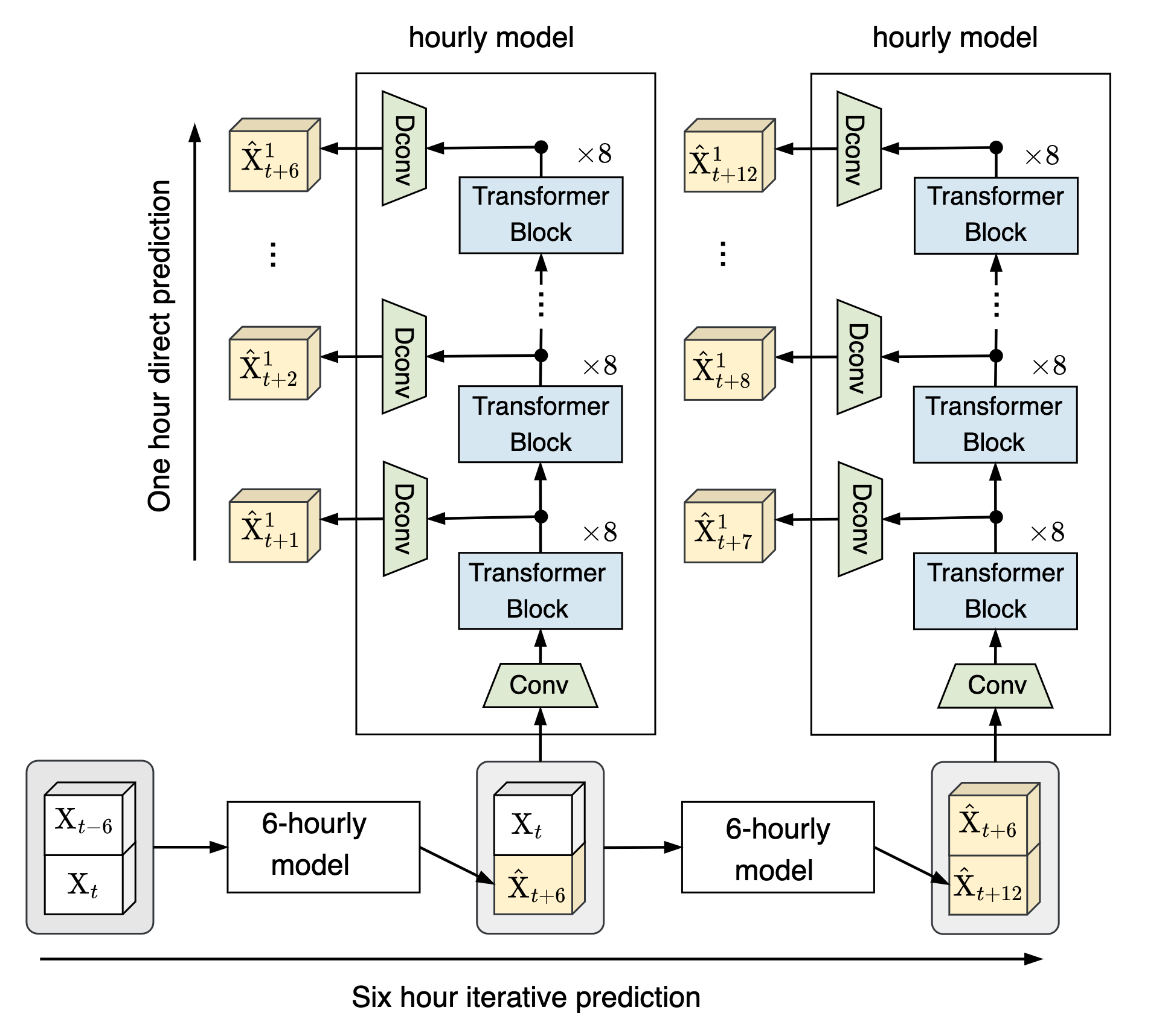}
    \caption{Schematic diagram of the structures of the FuXi-2.0 model.}
    \label{model}    
\end{figure}

The architecture of the 6-hourly model in FuXi-2.0 closely resembles that of the 1-hourly model illustrated in Figure \ref{model}.
Both models within FuXi-2.0 process meteorological data arranged in a four-dimensional cube with dimensions of $2\times88\times721\times1440$, where ${2}$ represents two prior time steps ($t-6$ and $t$ for 6-hourly model, and $t+6$ and $t$ for 1-hourly model) at 6-hour intervals, ${88}$ represents the total number of meteorological variables (${C}$), and ${721}$, and ${1440}$ denote the number of latitude (${H}$) and longitude (${W}$) grid points, respectively.
Initially, the data cube is reshaped to a dimension of $(2C)\times H \times W$ and then reduced to $2C\times90\times180$ using a 2-dimensional (2D) convolution layer with a kernel of 8 and a stride of 8.
In parallel, geographical and temporal variables are encoded and processed by a 2D convolution layer with the same kernel size and stride.
The encoded variables are then concatenated with the input meteorological data of reduced dimensions.
The concatenated data is subsequently processed through 30 consecutive Swin Transformer \cite{liu2021swin} blocks, each with a window size of $18 \times 18$.
This phase incorporates a Rotary Position Embedding (RoPE) \cite{su2024roformer,heo2024rotary}, a state-of-the-art position embedding technique that combines the advantages of both absolute and relative positional encodings.
The final step involves reconstructing the data to its original dimensions using a 2D transposed convolution layer with a kernel size and stride of 8 \cite{Zeiler2010}.

The 1-hourly forecast shares a similar structure with the 6-hourly model, as shown in Figure \ref{model}, including 30 Swin Transformer blocks.
A key difference in this model is the arrangement of 5-block subsets, each followed by a 2D transposed convolution layer with a kernel and stride of 8.
This configuration allows the 1-hourly forecast model to generate 6 consecutive hourly forecasts within a 6-hour window, significantly improving efficiency by reducing the iterations required compared to the Pangu-Weather model.
Additionally, this model enhances forecast accuracy by incorporating 6-hourly forecast data, effectively leveraging comprehensive data over these intervals. 
Unlike the Pangu-Weather model, which employs 4 separate models and varying iteration counts for hourly forecasts, the FuXi-2.0 model maintains a consistent iteration count.
For example, the Pangu-Weather model requires only 3 iterations to generate a 72-hour forecast but 8 iterations for a 71-hour forecast. In contrast, FuXi-2.0 requires 12 iterations for a 72-hour forecast and 13 iterations to generate hourly forecasts within the 6-hour interval between 66 and 72 hours.

\FloatBarrier

\subsection{FuXi-2.0 model training}

The loss function used is the robust Charbonnier loss \citep{Charbonnier1994}, computed between the model's output ( $\hat{\textrm{\textbf{X}}}^{t+1}$) and the target ${\textrm{\textbf{X}}}^{t+1}$, defined as:
\begin{equation}
\label{loss}
L = \sqrt{(a_i(\hat{\textrm{\textbf{X}}}^{t+1} - \textrm{\textbf{X}}^{t+1}))^2 + \epsilon^2}
\end{equation}
where ${\epsilon}$ is a constant set to ${10^{-3}}$ and $a_i$ denotes the weight at latitude ${i}$ which decreases as latitude increases. The Charbonnier loss is a mix between L1 and L2 loss. It functions like L1 regularization when the error exceeds epsilon, and it acts more like an L2 loss when the error falls below epsilon. The Charbonnier loss is favored over L1 and L2 losses due to its consistently superior performance. Additionally, different weights are assigned to the upper-air (1) and surface variables (0.1) within the loss function.
The loss is set to 0 for oceanic variables over land areas.

The FuXi-2.0 model is developed using the Pytorch framework \citep{Paszke2017}. The 6-hourly model of FuXi-2.0 model is trained with 40,000 iterations, employing a batch size of 1 on each GPU. The optimization utilized the AdamW optimizer \cite{kingma2017adam,loshchilov2017decoupled} with parameters ${\beta_{1}}$=0.9 and ${\beta_{2}}$=0.95, an initial learning rate of 2.5$\times$10$^{-4}$, and a weight decay coefficient of 0.1. Upon completion of training the 6-hourly model and fixing its model parameters, the 1-hourly model is trained with 60,000 iterations on a cluster of 8 Nvidia A100 GPUs. In training the 1-hourly model, the 6-hourly model is run for a single step, generating the model output ($\hat{\textrm{\textbf{X}}}^{t+6}$). Subsequently, $\textrm{\textbf{X}}^{t}$ and the model output $\hat{\textrm{\textbf{X}}}^{t+6}$ served as input for the 1-hourly model, producing the 1-hourly forecasts ($\hat{\textrm{\textbf{X}}}^{t+1}$, $\hat{\textrm{\textbf{X}}}^{t+2}$, $\hat{\textrm{\textbf{X}}}^{t+3}$, $\hat{\textrm{\textbf{X}}}^{t+4}$, $\hat{\textrm{\textbf{X}}}^{t+5}$, and $\hat{\textrm{\textbf{X}}}^{t+6}$).

\subsection{Wind power forecasting model}
\label{power_model}

The National Energy Bureau (NEB) mandates wind power forecasting across various timescales \cite{ChineseGBT}. This study focuses on evaluating day-ahead short-term wind power forecasts using FuXi-2.0 and ECMWF HRES, both initialized at 12 UTC, with lead times ranging from 28 to 51 hours. Although NEB guidelines \cite{ChineseGBT} specify a 15-minute temporal resolution for these forecasts, we simplify to 1-hour resolution for this analysis.

We employ a Multi-Layer Perceptron (MLP) model for wind power forecasting, which is more critical for wind farms than mere wind speed forecasts. While various ML models could be applied for this conversion, comparing these models is beyond the scope of this study.
The input features include wind speed and direction at both 100 and 10 meters: ${\textrm{U100M}}$, ${\textrm{V100M}}$, ${\textrm{WS100M}}$, ${\textrm{U10M}}$, ${\textrm{V10M}}$, and ${\textrm{WS10M}}$. To capture spatial and temporal dynamics, data from a 3 $\times$ 3 grid (0.25\textdegree) around the wind farms and time steps from 1 hour before to 1 hour after the target forecasting step are included. In the MLP model, all input variables ($162 = 6 \times 3 \times 3 \times 3$) are reshaped into a one-dimensional (1D) tensor and passed through 5 hidden layers comprising 288, 256, 128, 64, and 32 neurons, respectively, using the rectified linear unit (ReLu) \cite{agarap2018deep} activation function. The loss function used is the RMSE between the predicted and actual power output.
The models are trained for 50 epochs with a batch size of 64 on an NVIDIA A100 GPU, taking about 20 minutes per model.
All the models are developed using the Pytorch framework \citep{Paszke2017}. 
The optimization utilized the Adam optimizer \cite{kingma2017adam} with parameters ${\beta_{1}}$=0.9 and ${\beta_{2}}$=0.999.
an initial learning rate of 1$\times$10$^{-4}$, and a weight decay coefficient of 0.2.
The learning rate is initialized as 1$\times$10$^{-4}$ and decays by 0.2 once the number of epochs reaches one of the milestones (15, 20, 25, 30, 35, 40, 45) by using the StepLR scheduler, until the final learning rate equals to 1.28$\times$10$^{-9}$.


\subsection{Evaluation method}
\label{eval_method}

To ensure a fair comparison among models, it is essential to evaluate each model against its corresponding initial conditions. Specifically, when evaluating the performance of ECMWF HRES, the HRES-fc0 dataset, which is used to initialize ECMWF HRES forecasts, should be considered as the ground truth. However, the temporal resolution of HRES-fc0 data is limited to 6 hours due to the initialization of ECMWF HRES occurring four times daily (00/06/12/18 UTC). Therefore, for the evaluation of 1-hourly forecasts generated by all models, the ERA5 reanalysis data, available at 1-hour intervals, becomes the only viable reference for evaluation. In the supplementary materials, the HRES-fc0 dataset served as the reference for HRES for evaluating 6-hourly forecasts of ECMWF HRES. Meanwhile, ERA5 remains as the reference for evaluating ML models, such as FuXi-2.0 and Pangu-Weather, both of which are trained and initialized using ERA5.

We follow standard verification practice \cite{rasp2023weatherbench,benbouallegue2023rise} in evaluating deterministic forecasts using the widely-used metrics, such as root mean squared error (RMSE), anomaly correlation coefficient (ACC), and forecast/observation activity. The formal definitions of these evaluation metrics are:

\begin{equation}
    \textrm{RMSE}(c, \tau) =\frac{1}{\mid \textrm{D} \mid}\sum_{t_0 \in \textrm{D}} \sqrt{\frac{1}{\textrm{H} \times \textrm{W}} \sum_{i=1}^\textrm{H}\sum_{j=1}^{\textrm{W}} a_i {( \hat{\textrm{\textbf{X}}}^{t_0 +\tau}_{c,i,j} - \textrm{\textbf{X}}^{t_0 +\tau}_{c,i,j} )}^{2}}
\end{equation}

\begin{equation}
    \textrm{ACC}(c, \tau) = \frac{1}{\mid \textrm{D} \mid}\sum_{t_0 \in \textrm{D}} \frac{\sum_{i, j} a_i (\hat{\textrm{\textbf{X}}}^{t_0 +\tau}_{c,i,j} - \textrm{M}^{t_0 +\tau}_{c,i,j}) (\textrm{\textbf{X}}^{t_0 +\tau}_{c,i,j} - \textrm{M}^{t_0 +\tau}_{c,i,j})} {\sqrt{ \sum_{i, j} a_i (\hat{\textrm{\textbf{X}}}^{t_0 +\tau}_{c,i,j} - \textrm{M}^{t_0 +\tau}_{c,i,j})^2 \sum_{i, j} a_i(\textrm{\textbf{X}}^{t_0 +\tau}_{c,i,j} - \textrm{M}^{t_0 +\tau}_{c,i,j})^2}}
\end{equation}

\begin{multline}
  \widehat{\textrm{Act}}(c, \tau) =\frac{1}{\mid \textrm{D} \mid} \sum_{t_0 \in \textrm{D}} \\
  \sqrt{\frac{1}{\textrm{H} \times \textrm{W}} \sum_{i=1}^\textrm{H}\sum_{j=1}^{\textrm{W}} a_i {[  ( \hat{\textrm{\textbf{X}}}^{t_0 +\tau}_{c,i,j} - \textrm{M}^{t_0 +\tau}_{c,i,j} ) - \frac{1}{\textrm{H} \times \textrm{W}} \sum_{i=1}^\textrm{H}\sum_{j=1}^{\textrm{W}} a_i {( \hat{\textrm{\textbf{X}}}^{t_0 +\tau}_{c,i,j} - \textrm{M}^{t_0 +\tau}_{c,i,j} )} ]}}
\end{multline}


\begin{multline}
  \textrm{Act}(c, \tau) =\frac{1}{\mid \textrm{D} \mid} \sum_{t_0 \in \textrm{D}} \\ 
  \sqrt{\frac{1}{\textrm{H} \times \textrm{W}} \sum_{i=1}^\textrm{H}\sum_{j=1}^{\textrm{W}} a_i {[  ( \textrm{\textbf{X}}^{t_0 +\tau}_{c,i,j} - \textrm{M}^{t_0 +\tau}_{c,i,j} ) - \frac{1}{\textrm{H} \times \textrm{W}} \sum_{i=1}^\textrm{H}\sum_{j=1}^{\textrm{W}} a_i {( \textrm{\textbf{X}}^{t_0 +\tau}_{c,i,j} - \textrm{M}^{t_0 +\tau}_{c,i,j} )} ]}}
\end{multline}
where ${t_0}$ denotes the forecast initialization time in the testing dataset $\textrm{D}$, and ${\tau}$ represents the lead time steps added to ${t_0}$. The climatological mean, symbolized by $\textrm{M}$, is derived from the ERA5 reanalysis data spanning the years from 1993 to 2016, which provides insight into the average climatic conditions during this period. Additionally, the forecast/observation activity is defined as the standard deviation of the forecast/observation anomaly, which assesses the variability or smoothness in the forecast/observation. Specifically, lower values of forecast activity ($\widehat{\textrm{Act}}$) or observation activity ($\textrm{Act}$) suggests a smoother forecast or observation, respectively. This measure helps to evaluate the consistency and reliability of the forecast. To more effectively distinguish between models with marginal performance differences, we use the normalized RMSE difference between model A and the baseline model B, calculated using the formula \((\textrm{RMSE}_\textrm{A}-\textrm{RMSE}_\textrm{B})/\textrm{RMSE}_\textrm{B}\). Similarly, the normalized ACC and forecast activity differences are formulated as \((\textrm{ACC}_\textrm{A}-\textrm{ACC}_\textrm{B})/(1-\textrm{ACC}_\textrm{B})\) and \((\textrm{Act}_\textrm{A}-\textrm{Act}_\textrm{B})/\textrm{Act}_\textrm{B}\). Here, negative values in the normalized RMSE difference and positive values in the normalized ACC and activity differences suggest that model A outperforms the baseline model B. The baseline model for calculating the normalized RMSE and ACC differences is ECMWF HRES, while the reference for calculating the normalized activity difference is ERA5.

To evaluate and compare the performance of wind power forecasts using data from ECMWF HRES and FuXi-2.0, the RMSE and Pearson’s correlation coefficient (CC) are used. Additionally, the accuracy of day-ahead wind power forecast is calculated according to the following equation:
\begin{equation}
    \textrm{A}_{WP} = (1 - \frac{\sqrt{\sum_{t=1}^{n} (\textrm{P}_{Mt} - \textrm{P}_{Pt})^2}}{Cap \times \sqrt{n}}) \times 100%
\end{equation}
where $\textrm{P}_\textrm{Mi}$ and $\textrm{P}_\textrm{Pi}$ represent the actual and predicted power at time $t$, respectively, while $Cap$ is the rated capacity of wind farms. $n$ denotes the number of samples, which is 24 in this study. 


\section*{Data Availability Statement}
We downloaded a subset of the ERA5 dataset from the official website of Copernicus Climate Data (CDS) at \url{https://cds.climate.copernicus.eu/}. The ECMWF HRES forecasts are available at \url{https://apps.ecmwf.int/archive-catalogue/?type=fc&class=od&stream=oper&expver=1}.
The preprocessed sample data used for running FuXi-2.0 models in this work are available at \url{https://zenodo.org/records/11065422} \cite{code2024}. The wind power data for Kelmarsh wind farm is downloaded from \url{https://doi.org/10.5281/zenodo.5841834}.
Wind power data for Yeongheung was obtained from the Republic of Korea public data portal (\url{www.data.go.kr})


\section*{Code Availability Statement}
The Pangu-Weather model is available on GitHub (\url{https://github.com/198808xc/Pangu-Weather}).

\section*{Acknowledgements}
We express our sincere gratitude to the researchers at ECMWF for their invaluable contributions to collecting, archiving, disseminating, and maintaining the ERA5 reanalysis dataset and ECMWF HRES.

The computations in this research were performed using the CFFF platform of Fudan University.

\section*{Competing interests}
The authors declare no competing interests.


\noindent


\bibliography{refs}


\end{CJK*}

\clearpage

\appendix

\renewcommand\thefigure{\thesection.\arabic{figure}}    

\setcounter{figure}{0}    

\end{document}